\documentclass[preprint,showpacs,preprintnumbers,amsmath,amssymb]{revtex4}
\usepackage{graphicx}% Include figure files
\usepackage{dcolumn}% Align table columns on decimal point
\usepackage{bm}% bold math

\begin{document}

\title{Role of the Nuclear and Electromagnetic Interactions in the Coherent Dissociation of the Relativistic $^7$Li Nucleus into the $^3$H~+~$^4$He Channel}

\author{N.~G.~Peresadko}
   \affiliation{Lebedev Physical Institute, Russian Academy of Sciences, Leninski pr. 53, Moscow, 119991 Russia} 
\author{V.~N.~Fetisov}
   \email{fetisov@sci.lebedev.ru}
   \affiliation{Lebedev Physical Institute, Russian Academy of Sciences, Leninski pr. 53, Moscow, 119991 Russia} 
 \author{Yu.~A.~Aleksandrov}
   \affiliation{Lebedev Physical Institute, Russian Academy of Sciences, Leninski pr. 53, Moscow, 119991 Russia} 
\author{S.~G.~Gerasimov}
   \affiliation{Lebedev Physical Institute, Russian Academy of Sciences, Leninski pr. 53, Moscow, 119991 Russia}
\author{V.~A.~Dronov}
   \affiliation{Lebedev Physical Institute, Russian Academy of Sciences, Leninski pr. 53, Moscow, 119991 Russia}
\author{V.~G.~Larionova}
   \affiliation{Lebedev Physical Institute, Russian Academy of Sciences, Leninski pr. 53, Moscow, 119991 Russia}   
\author{E.~I.~Tamm}
   \affiliation{Lebedev Physical Institute, Russian Academy of Sciences, Leninski pr. 53, Moscow, 119991 Russia}
\author{S.~P.~Kharlamov}
   \affiliation{Lebedev Physical Institute, Russian Academy of Sciences, Leninski pr. 53, Moscow, 119991 Russia}

\begin{abstract}
\indent 
The differential cross section in the transverse momentum $Q$ and a total cross section of $(31\pm4)$~mb for the coherent dissociation of a 3-A-GeV/$c$~$^7$Li nucleus through the $^3$H$+^4$He channel have been measured on emulsion nuclei. The observed $Q$ dependence of the cross section is explained by the predominant supposition of the nuclear diffraction patterns on light (C, N, O) and heavy (Br, Ag) emulsion nuclei. The contributions to the cross section from nuclear diffraction ($Q\le400$~MeV/$c$) and Coulomb $(Q\le50$~MeV/$c$) dissociations are calculated to be 40.7 and 4~mb, respectively.\par
\indent \\
DOI: 10.1134/S0021364008140014\par
\end{abstract}
  %    {PACS-key}{21.45.+v} \and
   %   {PACS-key}{23.60+e} \and
    %  {PACS-key}{25.10.+s}  

\maketitle

\indent The properties of the nuclei and mechanisms of the reactions induced by the Coulomb and nuclear interactions in nucleus-nucleus collisions have been studied for more than five decades \cite{Landau,Serber,Alder,Sitenko}. These investigations have been recently expanded to the relativistic energy range \cite{Bertulani,Azhgirei,web,Adamovich}.\par

\indent It is known that the nuclear diffraction mechanism of the reactions at low momentum transfers $Q$ (similar to optical diffraction), which was predicted as early as the 1950s \cite{Akhiezer,Akhieser1,Glauber,Pomeranchuk}, becomes significant at energies of about hundreds of MeVs and higher along with the Coulomb interaction. Diffraction is characterized by the observed oscillations of the cross sections for the elastic scattering of particles and nuclei, $d\sigma/dQ$, with the main maximum at small angles $\vartheta\simeq\lambda/R$, where $\lambda$ is the de Broglie wavelength of the incident particle and $R$ is the radius of the nuclear interaction region. It is also known \cite{Pomeranchuk} that the diffraction mechanism can induce the coherent dissociation of the incident nucleus (without the excitation of the target nucleus) and the production of particles. The dissociation of the $^{12}$C nucleus into three $\alpha$ particles was observed at relativistic energies \cite{Engelage,Belaga}. However, the direct observation of the diffraction pattern with the counter technique (measurement of $d\sigma/dQ$) in the nuclear dissociation remains a sufficiently complex problem. The energy spectra of the charged particles at given angles \cite{Okamura,Heide} are normally used in such experiments.\par

\indent Among numerous reactions accompanying the collisions of relativistic nuclei (multifragmentation, meson production), we take a comparatively simple channel of coherent (elastic) dissociation of the $^7$Li nucleus ($\lambda\simeq0.01$~fm), which corresponds to the twocluster structure of the $^7$Li nucleus and is convenient for the application of the developed theoretical approaches to the description of such reactions.\par 	

\indent The cross section $d\sigma/dQ$ for the elastic dissociation of the $^7$Li nucleus is measured in the experiment in order to analyze the diffraction pattern of the process and to determine the contribution from the electromagnetic dissociation. According to previous nuclear emulsion measurements \cite{Adamovich}, the chosen reaction is characterized by very small nucleus emission angles and correspondingly low values $Q\le0.45$~GeV/$c$; under these conditions, the simultaneous manifestation of the Coulomb and nuclear diffraction mechanisms of the process can be expected \cite{Bertulani}. The $Q$ regions of the contributions from the Coulomb and nuclear interactions are revealed in theoretical approaches \cite{Bertulani,Bertulani1,Akhieser1,Davidovskii} with the two-cluster model of the $^7$Li nucleus with Pauli forbidden states \cite{Neudatchin,Kukulin}.\par

\indent An emulsion chamber composed of the BR-2 emulsion layers sensitive to the minimum ionization of single-charged particles was irradiated in a 3-A-GeV/$c~^7$Li beam from the JINR synchrophasotron. The tracks of the $^7$Li nuclei, as well as the single- and double-charged relativistic fragments, are specifically identified visually in terms of the ionization density. The fragment mass numbers $A_f$ are determined from the measurements of the average angles of the multiple Coulomb scattering of the fragments. The procedure of the identification of protons, deuterons, and $^3$H and $^{3,4}$He nuclei with the use of the momentum distributions of particles was described in detail in \cite{Adamovich}.\par

\indent Of a total number of 3730 inelastic interaction events, we select 85 events of the elastic dissociation of $^7$Li through the $^3$H$+^4$He channel, which are not accompanied by the destruction of the target nuclei and the emission of other particles. A nuclear emulsion contains $1.03\times10^{22}$~cm$^{-3}$ of Ag and Br nuclei and $2.85\times10^{22}$~cm$^{-3}$ of C, N, and O nuclei close in mass. The mean free paths of the Ag and Br recoil nuclei with momenta lower than 1~GeV/$c$ are so short that they are not detected. The C, N, and O nuclei with momenta higher than 200-300 MeV/$c$ have mean free paths longer than 2~$\mu$m and can be identified. In the selected events, $Q\le400$~MeV/$c$ and the recoil nuclei are not observed. The events of the interactions of $^7$Li with the emulsion protons were identified but are not considered in this work.\par

\indent These 85 events were identified when the beam tracks with a total length of 548.37 m were analyzed; the corresponding mean free path for this reaction is $6.5\pm0.7$~m. The reaction cross section averaged over all of the emulsion nuclei is determined as the ratio of the number of events to the total number of nuclei in the length of the analyzed tracks: $\sigma=85/5.4837\times10^4$~cm$\times4.91\times10^{22}$~cm$^{-3}=31\pm4$~mb.\par

\begin{figure}
    \includegraphics[width=4in]{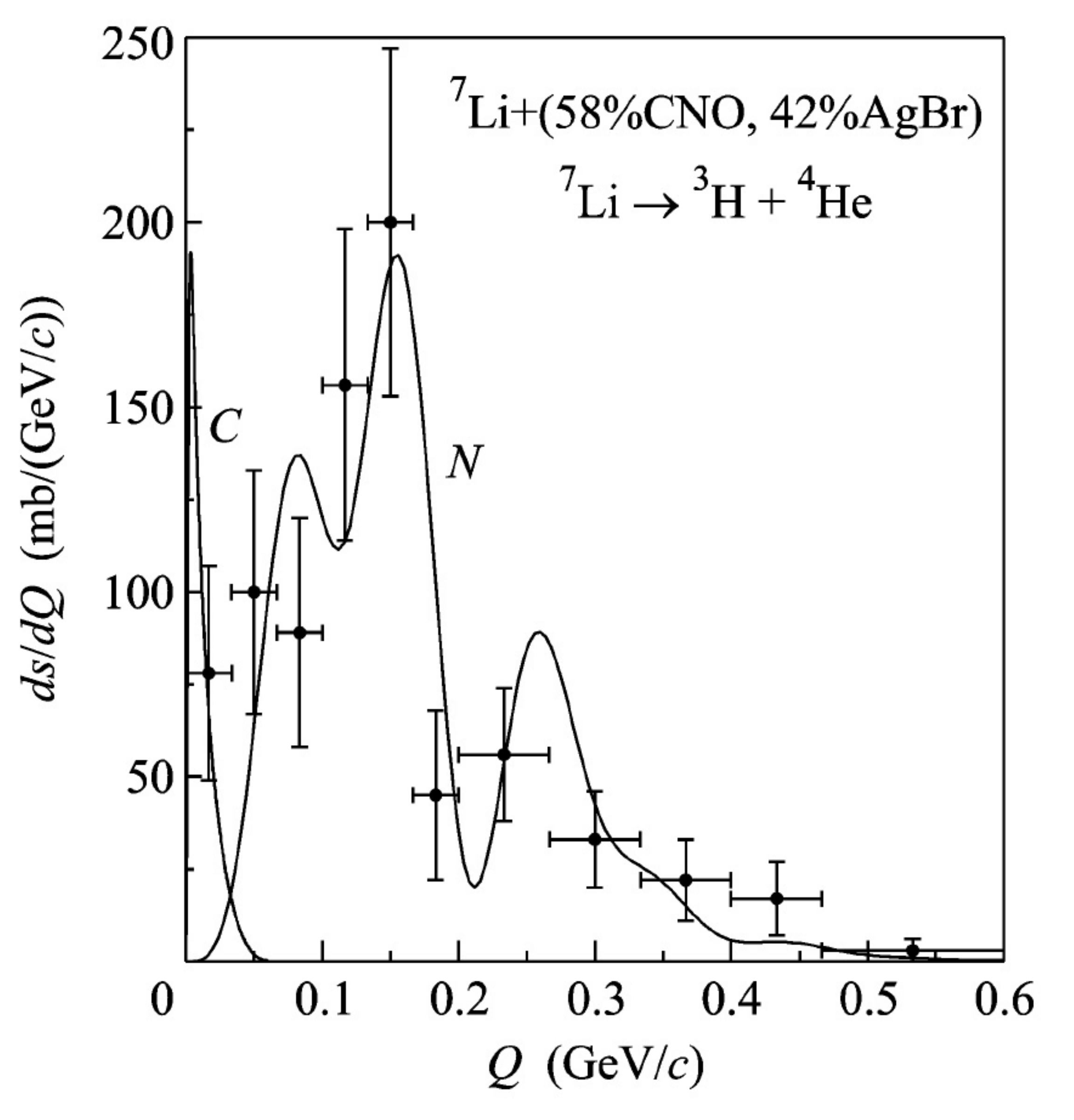}
    \caption{\label{Fig:1} Experimental and theoretical cross sections for ($C$) Coulomb and ($N$) nuclear diffraction dissociations of the $^7$Li nuclei.}
    \end{figure}
		
\indent The transverse momentum transfer $Q$ is the sum of the transverse momenta of $^3$H and $^4$He. Their magnitudes are given by the expression $p_t=p_0A_f\sin(\theta)$, where $\theta$ is the fragment emission angle with respect to the initial direction of the $^7$Li nucleus. The accuracy of measuring the $Q$ values is estimated as $\sim10$~MeV/$c$. The experimental cross section $d\sigma/dQ$ is shown in Fig. 1 along with the theoretical lines described below. The characteristic nonmonotonic $Q$ dependence of the cross section with a maximum in the region 100-170 MeV/$c$ and a minimum near 200 MeV/$c$ remains unchanged when the histogram step $\Delta Q$ varies from 15 to 40 MeV/$c$.\par

\indent The $^{6,7}$Li and $^7$Be nuclei are the lightest 1$p$ nuclei. As known, these nuclei have a two-cluster structure with a high probability ($0.8-1.0$) \cite{Neudatchin1}. In this work, the $^7$Li nucleus and the states of the $^3$H and $^4$He clusters in the continuous spectrum are described in the potential model with the Pauli forbidden states formulated in \cite{Neudatchin,Kukulin}. This model well describes the static properties of the light cluster nuclei, nuclear form factors, photodissociation processes, and scattering phases \cite{Dubovichenko,Burkova}.\par

\indent The cluster interaction potential is written as the sum of three terms corresponding to the central $V$, spin-orbit $V_{sc}$, and Coulomb $V_C$ interactions: \\
\begin{equation}
\begin{split}
V(r)=-V_0(1+exp[r-R_c)/a])^{-1},\\
V_{sc}(r)=-V_1\mathbf{ls}\frac{d}{dr}V(r)
\end{split}
\end{equation} 
\begin{equation}
\begin{split}
V_C(r)=\begin{cases}
\frac{Z_1Z_2e^2}{2R_c}\left(3-\frac{r^2}{R^2_c}\right),~~r\le R_c \\
\frac{Z_1Z_2e^2}{r},~~r>R_c
\end{cases}
\end{split}
\end{equation} \\
with the previously found parameters \cite{Neudatchin,Dubovichenko}:\\
\begin{equation}
\begin{split}
V_{00}=98.5~\mbox{MeV},~\Delta V=11.5~\mbox{MeV}, \\
R_c=1.8~\mbox{fm},~a=0.7~\mbox{fm}, \\
V_0=V_{00}+\Delta V(-1)^{l+1}, \\
V_1=0.015(3+(-1)^{l+1})~\mbox{fm}^2.
\end{split}
\end{equation} \\
To calculate the cross sections, we used the wavefunctions of two allowed bound states $P_{3/2}(-2.36$~MeV), which is the ground state of $^7$Li, and $P_{1/2}(-1.59$~MeV) and six forbidden bound states $S_{1/2}(-57.4$~MeV), $S_{1/2}(-15.9$~MeV), $P_{3/2}(-34.4$~MeV), $P_{1/2}(-32.3$~MeV), $D_{5/2}(-13.7$~MeV), and $D_{3/2}(-11.1$~MeV) in this potential. \par

\indent An important assumption of the developed theories of the Coulomb dissociation of relativistic nuclei in quantum \cite{Bertulani1} and semiclassical \cite{Winther} approaches is that the Coulomb amplitude is much smaller than the nuclear amplitude for impact parameters $b\le R$, where $R$ is about the sum of the radii of the colliding nuclei. It is accepted that the nuclear dissociation mechanism dominates for such $b$ values.\par

\indent The cross section for the Coulomb dissociation of $^7$Li is calculated in the Bertulani-Baur formalism \cite{Bertulani,Bertulani1} with the use of the multipole expansion of the electromagnetic interaction. The main contributions to the cross section come from the $E$1 transitions $P_{3/2}\rightarrow S_{1/2},~D_{3/2},~D_{5/2}$. Performing the integration with respect to the angles of the momenta of emitted clusters in the initial expression for the cross section \cite{Bertulani,Bertulani1}, we arrive at the expression \\
\begin{equation}
\begin{split}
\frac{d\sigma_C}{dQ}=\frac{32}{9}\left(\frac{Ze^2}{\hslash\textsl{v}}\right)^2c_dQR^2\int\limits_0^{\infty}\frac{\xi^2}{(\xi^2+(QR)^2)^2}\left(I_2^2(k)+\frac{1}{2}I^2_{0,1/2}(k)\right)\left(f^2_1+\frac{1}{\gamma^2}f^2_0\right)k^2dk.
\end{split}
\end{equation} \\
Here, the functions $f_n$ and radial integrals $I_{l,j}(k)$ for the dipole transitions are given by the respective expressions \\
\begin{equation}
\begin{split}
f_n=\xi J_n(QR)K_{n+1}(\xi)-QRJ_{n+1}(QR)K_n(\xi),\\
I_{l,j}(k)=\int\limits_0^{\infty}R_{l,j}(k,r)R_i(r)r^3dr,
\end{split}
\end{equation} \\
where $J_n$ and $K_n$ are the Bessel functions; $l$ and $j$ are the orbital and total angular momenta, respectively; and $R_i$ and $R_{l,j}$ are the wavefunctions of the clusters in the ground state (binding energy is $E^{\mbox{exp}}_b=2.47$~MeV) and in continuum, respectively. The small difference between the $D_{3/2}$- and $D_{5/2}$-states is neglected in the integrals $I_{2,j}=I_2$. In the absence of an interaction, the functions $R_{l,j}(k, r)$ coincide with the spherical Bessel functions $j_l(kr)$. In Eq. (4), $Z$ is the charge number of the target nucleus; $\textsl{v}$ is the velocity of $^7$Li; the coefficient $c_d=(Z_1\beta_1-Z_2\beta_2)^2$, where $\beta_{1(2)}=m_{2(1)}/(m_1+m_2)$ and $m_i$ are the cluster masses, specifies the dipole moment of the cluster system; $\gamma=(1-(\textsl{v}/c)^2)^{-1/2}$; and $\xi=(\omega R)/(\gamma\textsl{v})$, where $\omega=E_b+(\hslash k)^2/2\mu_{\alpha t}$. The calculations are performed with the average values $\bar R=5.0$~fm and $\bar Z=7$ for the C, N, and O nuclei and $\bar R=8.1$~fm and $\bar Z=41$ for the Ag and Br nuclei. Line $C$ in Fig. 1 is the $Q$ dependence of the cross section in a very narrow interval $Q\le50$~MeV/$c$ with a maximum at $Q\simeq3.5$~MeV/$c$. The total cross section $\sigma_C$ calculated for the emulsion containing 58\% of the C, N, and O nuclei and 42\% of the Ag and Br nuclei is 4 mb. The interval $Q\le30$~MeV/$c$ contains only six events; the corresponding estimate of the cross section for Ag and Br nuclei is $\sigma_C=6/(5.4837\times10^4\mbox{cm}\times2.06\times10^{22}\mbox{cm}^{-3})=(5\pm2)$~mb, which is somewhat smaller than the value of 9.1~mb calculated for these nuclei. Since the $\sigma_C$ value and $Q$ interval are small, it is necessary to take into account the contribution from the nuclear fragmentation. \par

\begin{figure}
    \includegraphics[width=4in]{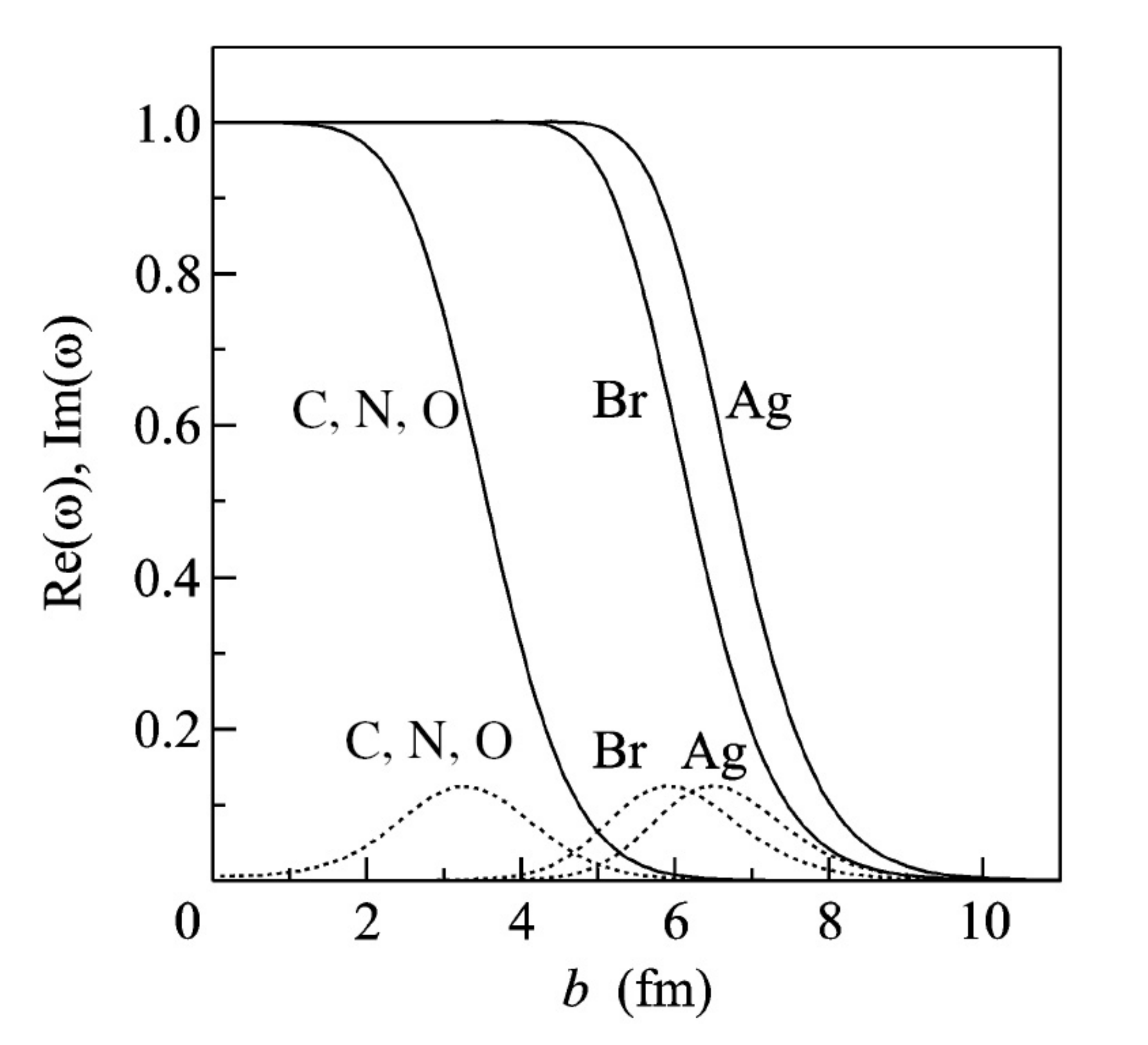}
    \caption{\label{Fig:2} (Solid lines) Real and (dotted lines) imaginary parts of the profile functions.}
    \end{figure}
	
\indent Following the Akhiezer-Sitenko formalism \cite{Akhieser1}, which was recently developed in \cite{Davidovskii} when applied to the diffraction scattering of two-cluster nuclei, the process cross section is determined by the matrix elements of the following combination of the profile functions $\omega(b)$: \\
\begin{equation}
\begin{split}
\omega_{\alpha}(b_{\alpha})+\omega_t(b_t)-\omega_{\alpha}(b_{\alpha})\omega_t(b_t),\\
\omega_i(b)=1-\exp(i\chi_i(b)).
\end{split}
\end{equation} \\
The phase functions $\chi_i$ describing the collisions between the nuclei with the mass numbers $A_1$ and $A_2$ are calculated in the optical limit of the Glauber-Sitenko model with the use of the formula for the convolution of the nuclear form factors $S_{A_i}(q)$ and $NN$ amplitude \cite{Franco}:\\
\begin{equation}
\begin{split}
i\chi(b)=-\frac{A_1A_2\sigma_N}{8\pi^2}(1-i\rho)\times\int\exp(-i\mathbf{qb}-a_nq^2/2)K(q)S_{A_1}S_{A_2}d^2q.
\end{split}
\end{equation} \\
The form factors for the $\alpha$ and $t$ clusters and C, N, and O nuclei are calculated in the oscillatory shell model with the correction $K(q)$ on the motion of the center of mass. The Fermi density distribution is taken for the Ag and Br nuclei. The parameters of the oscillatory model and Fermi distribution are fit in the standard way \cite{Tassie}, \cite{Franco1} with the experimental rms nuclear radii \cite{Barrett} $\bar r_t=1.7$~fm, $\bar r_{\alpha}=1.67$~fm, $\bar r_{C,N,O}=2.54$~fm, $\bar r_{Br}=5.1$~fm, and $\bar r_{Ag}=5.62$~fm. The accepted parameters of the $NN$ interaction are $\sigma N=43.0$~mb, $\rho=-0.35$, and $a_N=0.242$~fm$^2$ \cite{Franco1}. Since $\omega_{\alpha}$ and $\omega_t$ are close to each other, we use the half-sum of these values as $\omega(b)$ for the chosen target nucleus. The real and imaginary parts of $\omega(b)$ for light and heavy nuclei of the emulsion are shown in Fig. 2. The first two terms in Eq. (6), which correspond to the impulse approximation, make the main contribution to the cross section:\\
\begin{equation}
\begin{split}
\frac{d\sigma_N}{dQ}\!\!=\!\! A\!\!\left(\!\!1+I_0(Q)\!\!-\!\!\frac{3}{2}\sum_{lj,L}I^{lj}_L((\beta_1Q)+(-1)^LI^{lj}_L(\beta_2Q))^2\hat l\hat j(10l0|L0)^2\!\!\left\{\begin{array}{ccc}\!\! j & l & 1/2\!\! \\\!\!1 & 3/2 & L \end{array}\right\}^2\right)\!\!,
\end{split}
\end{equation} \\
\begin{equation}
\begin{split}
\frac{A}{4\pi Q}=\left|\int\limits_0^{\infty}\omega(b)J_0(Qb)bdb\right|^2,\\
I_0(q)=\int\limits_0^{\infty}j_0(qr)R^2_ir^2dr,\\
I^{lj}_L(q)=\int\limits_0^{\infty}j_L(qr)R_{lj}R_ir^2dr.
\end{split}
\end{equation} \\
Expression (8) was obtained taking into account the completeness of the states of the cluster Hamiltonian. This completeness makes it possible to exclude the integration over the states of the continuum and to express the cross section in terms of the matrix elements between all of the bound states $(l,j)$. In Eq. (8), $\hat l\hat j=(2l+1)(2j+1)$ and the next two factors are the squares of the Clebsch-Gordan coefficient and the $6j$ symbol. The calculations for nuclei with the sharp edge show that the alternating contribution to the cross section from the third term in Eq. (6), which corresponds to the simultaneous collision of two clusters with the target nucleus (eclipse term), is no more than 1-2\%. \par

\indent The cross sections $d\sigma_N/dQ$ for light (C, N, O, $\sigma_N=31.6$ mb) and heavy (Br, $\sigma_N=50.6$~mb; Ag, $\sigma_N=56.0$~mb) nuclei concentrate in the region $Q\le0.4$~GeV/$c$ and have the pronounced oscillating form with the oscillation period in $Q$ close to the zeros of the function $J_1(QR)$, where $R$ is about the size of the region of the profile function Re($\omega$). The imaginary parts Im($\omega$) make small contributions to the cross sections and lead to a small filling of the minima. Note that the cross section for a black nucleus with a sharp boundary is proportional to $J^2_1(QR)$~\cite{Akhieser1}. This strong-absorption model contradicts the experiment, because it gives many oscillations in a very wide diffraction cone $(Q\le2$~GeV/$c$) and provides a too large cross section (more than 200~mb). For nuclei with smeared surfaces, the theory accepted in this work predicts two maxima in the cross section for C, N, and O nuclei at $Q\simeq120$ and 280~MeV/$c$ with the intensity ratio 1 : 0.34 and four maxima for the Br and Ag nuclei at $Q\simeq70$, 170, 270, and 360~MeV/$c$ with the approximate intensity ratio 0.7 : 1.0 : 0.5 : 0.15. This is attributed to a large difference between the radii of the light and heavy nuclei and between the corresponding profile functions. Such a pattern of the inelastic coherent diffraction differs from the diffraction pattern in the elastic scattering, where the forward scattering peak dominates.\par

\indent The resulting cross section for the nuclear composition of the emulsion is shown by line $N$ in Fig. 1, where the experimental data are also presented. For the normalization to the total measured cross section, the nuclear and electromagnetic theoretical cross sections are multiplied by a common factor of $k=0.7$. As seen in the figure, the $Q$ dependence of the cross section is explained by the imposition of two diffraction patterns (individual oscillations of the cross sections) for the light and heavy nuclei of the emulsion. The regions of the Coulomb and nuclear dissociation mechanisms are well separated, and their interference, which is not considered in this work, is expected in the narrow range $0\le Q\le50$~MeV/$c$. Note that the theoretical total cross section $\sigma_N+\sigma_C=44.7$~mb is larger than the experimental value of $(31\pm4)$~mb. The difference between the cross sections is apparently caused by the use of the planewave impulse approximation \cite{Sitenko}. In addition, the $(\alpha,~t)$ clustering in $^7$Li is possibly incomplete.\par

\indent The results show that the two-cluster nucleus $^7$Li, as well as the deuteron, can be used as a probe nucleus to verify the theory of the electromagnetic and diffraction dissociations and to acquire information on the surface layer of the nuclei. It is interesting to perform experiments for the observation of the characteristic diffraction patterns of the coherent dissociation of $^7$Li into the $(\alpha,~t)$ channel on pure target nuclei in a wide range of mass numbers with the use of the counter technique.\par

\indent We are grateful to A.V. Pisetskaya and L.N. Shesterkina for their tremendous work on the search for and measurements of the events on microscopes, as well as to Prof. V.G. Neudatchin for valuable information on the theoretical investigations on the cluster structure of light nuclei at the Skobel'tsyn Institute of Nuclear Physics, Moscow State University, and stimulating discussions of the results. This work was supported by the Russian Foundation for Basic Research (project no. 07-02-00871-a).\par


\newpage


\begin{thebibliography}{}
\bibitem{Landau}
L. D. Landau, Phys. Zs. Sovietun. 1, 88 (1932); Collected Works, Vol. 6.
\bibitem{Serber}
R. Serber, Phys. Rev. 72, 1008 (1947).
\bibitem{Alder} 
K. Alder and A. Winther, Electromagnetic Excitation (North-Holland, Amsterdam, 1975).
\bibitem{Sitenko} 
A. G. Sitenko, Theory of Nuclear Reactions (Energoatomizdat, Moscow, 1983; World Sci., Singapore, 1990).
\bibitem{Bertulani} 
C. A. Bertulani and G. Baur, Phys. Rep. 163, 299 (1988).
\bibitem{Azhgirei} 
L. S. Azhgirei and N. P. Yudin, Fiz. Elem. Chastits At. Yadra 37, 1012 (2006) [Phys. Part. Nucl. 37, 535 (2006)].
\bibitem{web} 
The BECQUEREL Project, http://becquerel.jinr.ru/.
\bibitem{Adamovich} 
M. I. Adamovich et al., J. Phys. G 30, 1479 (2004).
\bibitem{Akhiezer} 
A. I. Akhiezer and A. G. Sitenko, Uch. Zap. Khark. Univ. 64, 9 (1955).
\bibitem{Akhieser1} 
A. I. Akhieser and A. G. Sitenko, Phys. Rev. 106, 1236 (1957).
\bibitem{Glauber} 
R. Glauber, Phys. Rev. 99, 1515 (1955).
\bibitem{Pomeranchuk} 
I. Ya. Pomeranchuk and E. L. Feinberg, Dokl. Akad. Nauk SSSR 93, 439 (1953); E. L. Feinberg, Zh. Eksp. Teor. Fiz. 29, 115 (1955) [Sov. Phys. JETP 2, 58 (1955)]; E. L. Feinberg and I. Ya. Pomeranchuk, Suppl. Nuovo Cim. 3, 652 (1956).
\bibitem{Engelage} 
J. Engelage et al., Phys. Lett. B 173, 34 (1986); F. A. Cucinotta and R. R. Dubey, Phys. Rev. C 50, 1090 (1994).
\bibitem{Belaga} 
V. V. Belaga, A. A. Bendzhaza, V. V. Rusakova, et al., Yad. Fiz. 58, 2014 (1993) [Phys. At. Nucl. 58, 1905 (1993)].
\bibitem{Okamura} 
H. Okamura et al., Phys. Rev. C 58, 2180 (1998).
\bibitem{Heide} 
N. Heide, D. K. Srivastava, and H. Rebel, Phys. Rev. Lett. 63, 601 (1985).
\bibitem{Bertulani1} 
C. A. Bertulani and G. Baur, Nucl. Phys. A 442, 739 (1985).
\bibitem{Davidovskii} 
V. V. Davidovskii, M. V. Evlanov, and V. K. Tartakovskii, Yad. Fiz. 69, 252 (2006) [Phys. At. Nucl. 69, 230 (2006)].
\bibitem{Neudatchin} 
V. G. Neudatchin and Yu. F. Smirnov, Current Problems in Optics and Atomic Physics (Kiev, 1974), p. 225 [in Russian].
\bibitem{Kukulin} 
V. I. Kukulin, V. G. Neudatchin, and Yu. F. Smirnov, Fiz. Elem. Chastits At. Yadra 10, 1236 (1979) [Sov. Phys. Part. Nucl. 10, 492 (1979)].
\bibitem{Neudatchin1} 
V. G. Neudatchin and Yu. F. Smirnov, Nucleon Associations in Light Nuclei (Nauka, Moscow, 1969) [in Russian].
\bibitem{Dubovichenko} 
S. B. Dubovichenko, The Properties of Ligth Atomic Nuclei in Potential Cluster Models (Denker, Almaty, 2004) [in Russian].
\bibitem{Burkova} 
N. A. Burkova, K. A. Zhaksybekova, and M. A. Zhusupov, Fiz. Elem. Chastits At. Yadra 36, 821 (2005) [Phys. Part. Nucl. 36, 427 (2005)].
\bibitem{Dubovichenko1} 
S. B. Dubovichenko and M. A. Zhusupov, Izv. Akad. Nauk KazSSR, Ser. Mat. Fiz., No. 4, 44 (1983).
\bibitem{Winther} 
A. Winther and K. Alder, Nucl. Phys. A 319, 518 (1979).
\bibitem{Franco} 
V. Franco and A. Tekou, Phys. Rev. C 16, 658 (1977).
\bibitem{Tassie} 
L. J. Tassie and F. C. Barker, Phys. Rev. 111, 940 (1958).
\bibitem{Franco1} 
V. Franco, Phys. Rev. 6, 748 (1972).
\bibitem{Barrett} 
R. C. Barrett and D. F. Jackson, Nuclear Sizes and Structure (Clarendon, Oxford, 1977).
\end{thebibliography}
\end{document}